\documentclass[%
 reprint,
nofootinbib,
 amsmath,amssymb,
 aps,
 prx,
 longbibliography,
 showkeys,
floatfix,
]{revtex4-2}

\usepackage{algpseudocode}
\usepackage{float}
\makeatletter
\let\float@end\float@end@ltx
\let\newfloat\newfloat@ltx
\makeatother
\usepackage{algorithm}
\makeatletter
\def\fnum@algorithm{\fname@algorithm~\thealgorithm}
\makeatother
\usepackage[utf8]{inputenc}
\usepackage[T1]{fontenc}
\usepackage{hyperref}
\usepackage{url}
\usepackage{booktabs}
\usepackage{amsfonts}
\usepackage{nicefrac}
\usepackage{microtype}
\usepackage{xcolor}
\usepackage{graphicx}
\usepackage{amsthm}
\usepackage[nameinlink,capitalise,noabbrev]{cleveref}
\usepackage{caption,placeins}
\newcommand{\panellabel}[1]{\par\smallskip{\centering\textbf{\textsf{#1}}\par}}
\usepackage{import}
\usepackage{tikz}
\usepackage{mathtools}
\usepackage{bbm}
\usetikzlibrary{positioning}

\hypersetup{
  hidelinks,
}

\newcommand{\erdosrenyi}{Erd\H{o}s--R\'enyi}

\newcommand{\bflambda}{\boldsymbol{\lambda}}

\newcommand{\CC}{\texorpdfstring{\textsc{cc}}{CC}}
\newcommand{\RACC}{\texorpdfstring{\textsc{racc}}{RACC}}
\newcommand{\SC}{\texorpdfstring{\textsc{sc}}{SC}}
\newcommand{\MRSC}{\texorpdfstring{\textsc{mrsc}}{MRSC}}
\newcommand{\LM}{\texorpdfstring{\textsc{lm}}{LM}}
\newcommand{\CWN}{\texorpdfstring{\textsc{cwn}}{CWN}}
\newcommand{\PPGN}{\texorpdfstring{\textsc{ppgn}}{PPGN}}
\newcommand{\GNN}{\texorpdfstring{\textsc{gnn}}{GNN}}
\newcommand{\RND}{\texorpdfstring{\textsc{rnd}}{RND}}
\newcommand{\PROTEINS}{\texorpdfstring{\textsc{proteins}}{PROTEINS}}
\newcommand{\NCI}{\texorpdfstring{\textsc{nci109}}{NCI109}}

\DeclareMathOperator{\lca}{lca}
\DeclareMathOperator{\depth}{depth}

\newcommand{\repolink}{\href{https://github.com/josefhoppe/random-abstract-cell-complexes/}{\texttt{github.com/\allowbreak{}josefhoppe/random-\allowbreak{}abstract-\allowbreak{}cell-complexes}}}
\newcommand{\cwnrepolink}{\href{https://github.com/josefhoppe/cwn-random-ccs/}{\texttt{github.com/\allowbreak{}josefhoppe/\allowbreak{}cwn-\allowbreak{}random-\allowbreak{}ccs}}}

\date{\today}

\begin{document}

\title{Random Abstract Cell Complexes}
\author{Josef Hoppe}
\author{Michael T. Schaub}%
 \email{\{hoppe, schaub\}@netsci.rwth-aachen.de}
\affiliation{%
 RWTH Aachen University
}%

\begin{abstract}
    In machine learning and signal processing, methods based on (abstract) cell complexes (\CC{}s) are being applied successfully in different applications.
However, there is currently a lack of random models, which are necessary for a thorough evaluation of these methods, both by generating synthetic data and as null models to compare results to.
We fill this gap by defining a random model for abstract cell complexes based on independent inclusion probabilities, similar to the \erdosrenyi{} model for graphs.
We also discuss how modifications to bias the model to incorporate more structure can be implemented.
The model can also be used to lift a graph (or other cell complex) to a cell complex of arbitrary dimension.
The main challenge in this is that the number of cycles (possible 2-cells) on graphs is exponential, and the same is true in higher dimensions.
We overcome this complexity with an approximate sampling algorithm limited to 2-cells (polygons) that is based on a recent advance in sampling cycles on a graph.
Using this algorithm, we analyze the properties of random $2$-dimensional \CC{}s and showcase its usage as a performance baseline for both topological signal processing and topological deep learning.
The algorithm is available in the package \texttt{py-raccoon} on the Python Package Index.

\end{abstract}


\maketitle

\begin{figure*}
  \centering\includegraphics[width=\linewidth]{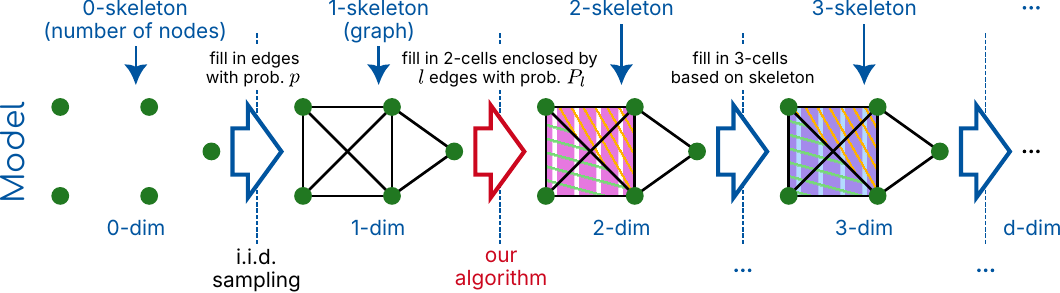}
  \caption{Overview of the random abstract cell complex model. The model is defined through iterative liftings from $k$- to $(k+1)$-dimensional cell complexes. This makes it possible to fix a skeleton of the complex and only apply the random lifting for subsequent dimensions. In this paper, we focus on the lifting from one- to two-dimensional abstract cell complexes. For this, we present a novel, efficient approximate sampling algorithm (see \cref{fig:sampling}).}\label{fig:model}
\end{figure*}

\section{Introduction}
Since the seminal work of Erd\H{o}s and R\'enyi \cite{erdos1959random}, random models of relational data described by graphs have been a mainstay topic of interest within probability theory, statistics, data science, and machine learning.
We motivate the interest in such random graphs from three perspectives.
First, they are vital as generative models for complex networks in various disciplines, including social sciences, physics, and biology~\cite{albert2002statistical,drobyshevskiy2019random}, as they enable the theoretical and empirical study of how certain (random) connection mechanisms influence the structure and properties of a network.
Second, random models can serve as null models to evaluate whether some observed network statistics may be deemed significant.
Third, random models can generate synthetic input data for training machine learning models, or for controlled experiments and benchmarking of network analysis algorithms.

Starting from the \erdosrenyi{} model~\cite{erdos1959random} and its variation by Gilbert~\cite{gilbert1959random}, a vast array of models for random graphs has been developed, such as the Stochastic Block Model~\cite{holland1983stochastic}, the Configuration Model~\cite{fosdick2018configuring}, etc. (see \cref{app:relwork} for more related literature).
Many of these models allow for the direct or indirect specification of certain network statistics --- e.g., density, triangle counts, degree distribution --- to be preserved (either exactly or in expectation) in samples  drawn from the model, with all other charecteristics being maximally random.

Recently, there has been a trend toward studying higher-order network models such as hypergraphs and simplicial and cellular complexes~\cite{bick2023higher,torres2021and,battiston2020networks}.
These richer models for relational data have gained interest as they enable the study of complex network structures with interactions beyond pairwise relationships, which have been deemed to be essential to model certain systems.
In particular, (abstract) cell complexes feature prominently in topological data analysis~\cite{wasserman2018topological} and in the emerging areas of topological signal processing~\cite{barbarossa2020topological,schaub2021signal} and topological deep learning~\cite{hajij2022topological}.
In contrast to random graph models, the landscape of models for these higher-order networks is underdeveloped.
Currently, the only practically usable models for random abstract cell complexes with arbitrary polygons as $2$-cells are ad hoc models \cite{hoppe2023representing} without intentionally set non-uniform inclusion probabilities.
\Cref{app:relwork} provides a more in-depth overview of related work.

Sampling from a well-defined model has to overcome the size and complexity of the cycle space of the graph.
Counting cycles alone is \#P-hard \cite{valiant1979complexity}, and estimators for cycle statistics have a runtime that is exponential in the length of cycles considered (e.g., \cite{giscard2017evaluating,alon2008biomolecular}).
However, recently, it was shown that spanning-tree-based estimators for cycle statistics \cite{hoppe2026cycles} can efficiently sample cycles of length up to order at least $\sqrt{n}$ (on common graph types), with a runtime of $\mathcal O(m \cdot n^3)$, or $\mathcal O(m^{1+\epsilon})$ per spanning tree when using an efficient approximation.

\paragraph*{Contributions}
We define a general model for random abstract cell complexes that generalizes the Linial--Meshulam (\LM{}) and multi-parameter random simplicial complex (\MRSC{}) models~\cite{linial2006homological,costa2016random} for simplicial complexes (\SC{}s), but it may also be used to lift a graph to a  cell complex.
Based on recent advances in sampling from the set of cycles on a graph, we provide an algorithm that lifts graphs to $2$-dimensional cell complexes by randomly adding cells.
The algorithm also provides a way to sample $2$-dimensional cell complexes with a certain \emph{number} of $2$-cells rather than a fixed probability, which is often desirable in practice.

\paragraph*{Outline} In the remainder of this paper, we first introduce relevant concepts and notation in \Cref{sec:bg-notation} before presenting our random cell complex model in \Cref{sec:model}.
We introduce the novel sampling algorithm in \cref{sec:cell-sampling-alg}.
Finally, we investigate properties of random cell complexes in \cref{sec:model-properties} and provide case studies of applications of random \CC{}s in \cref{sec:applications}.

\section{Background and Notation} \label{sec:bg-notation}

\subsection{Graphs and Simple Cycles}
We consider undirected graphs $G=(V,E)$ with $n:=|V|$ nodes and $m:=|E|$ edges.
An \erdosrenyi{} random graph $G(n,p)$ has $n$ nodes and for any node pair $u,v \in V$, the edge $(u,v)$ exists independently with probability $p$.
A cycle $c$ is a closed simple path on $G$; its length $|c|$ equals the number of nodes (or edges) in $c$.
We denote the set of all cycles in $G$ by $C$.
We use $C_l$ to denote the set of all cycles $c\in C$ with length $|c|=l$.
If node $u$ or edge $(u,v)$ is part of the cycle $c$, we write $u \in c$ or $(u,v) \in c$, respectively.
We assume, without loss of generality,\footnote{On a graph with multiple connected components, we can consider each component separately.} that all graphs are connected.

\subsection{Abstract Regular Cell Complexes}
In general, abstract (regular) cell complexes are a generalization of graphs, adding higher-dimensional structure (cells).
From here on, we will use the term \emph{cell complex} to refer to \emph{abstract regular} cell complexes.
A cell complex $\mathcal{C}$ consists of sets $\mathcal{C}_k$ of $k$-dimensional cells (or $k$-cells).
$k$-cells are simple $k$-dimensional volumes, for example $2$-cells are polygons, $3$-cells are polyhedra, etc.
Cells are attached to their boundary, a $k$-cell is enclosed by a set of $(k-1)$-cells forming a simple cycle, a sphere, etc.
All cells in the boundary (e.g. all edges enclosing a polygon) are also part of the cell complex $\mathcal C$.
We call a cell $c_{k-1}$ that is in the boundary of $c_{k}$ a \emph{face} of $c_k$, and $c_k$ is a \emph{coface} of $c_{k-1}$.
Simplicial complexes are a special case of cell complexes in which every cell is also a simplex, i.e., all polygons are triangles, all polyhedra are tetrahedra.

While the model introduced in this paper (cf. \cref{fig:model}) considers cell complexes of arbitrary dimension, our sampling algorithm is limited to two-dimensional cell complexes.
For the purpose of sampling, we may think of a $2$-dimensional cell complex as a graph with additional $2$-cells (polygons) that are attached to simple cycles on the graph.
For a more formal and complete introduction, we refer the reader to \cite{hoppe2025cellcomplexes}.

\subsection{Boundaries and Hodge Laplacians}
This subsection is based on \cite{hoppe2025cellcomplexes}.
The core structure of \CC{}s is described by boundary matrices (signed incidence matrices).
The first boundary matrix $B_1 \in \{0, \pm 1\}^{\mathcal C_0 \times \mathcal C_1}$ connects edges to vertices.
Each column corresponds to an edge and has one entry with value $+1$ and one with value $-1$, defining the reference orientation of that edge.
The second boundary matrix $B_2 \in \{0, \pm 1\}^{\mathcal C_1 \times \mathcal C_2}$ defines the incidence between polygons and edges.
The signs are chosen such that $B_1B_2 = 0$, which orients the edges in the boundary consistently.

The boundaries determine the (reference) orientation of a cell.
This orientation is arbitrary, and can be flipped by multiplying the corresponding rows and columns in the boundary matrices with $-1$.
We say that two \CC{}s $\mathcal C$ and $\mathcal C'$ are equivalent if they are the same up to orientation flips.

From these boundaries, we can generalize the graph Laplacian $L=D-A=B_1B_1^\top$ to the $k$-th Hodge Laplacian $L_k = B_k^\top B_k + B_{k+1}B_{k+1}^\top$.
The Betti number $\beta_k = \dim \ker L_k$ characterizes the holes (dimension of the $k$-th (co-)homology) of the \CC{}.
In our practical examples, the $1$-skeleton (nodes and edges) is fixed, while the $2$-cells are sampled or inferred, depending on the example.
Therefore, we are mostly interested in the 1st up-Laplacian $L_1^{\mathrm{up}} = B_2 B_2^\top$.
Eigenvalues and eigenvectors of Hodge Laplacians are relevant for applications in topological signal processing.
We will use the biggest eigenvalue $\lambda_{\max}$ of the up Laplacian as an indicator because it reflects the connectivity of the cycle space.

\subsection{Orientability of \CC{}s}

To characterize the structure of a \CC{}, we also look at its orientability.
Intuitively, a \CC{} is orientable if and only if it has a consistent `upper' and `lower' or `inner' and `outer' side.
Formally, a cell complex $\mathcal C$ is called \emph{orientable} if there is an equivalent \CC{} $\mathcal C'$ that is consistently oriented.
A cell complex is consistently oriented if every boundary matrix $B_k$ of dimension $k \geq 2$ has at most one entry $+1$ and one entry $-1$ in each row.

What does orientatility mean in practice?
As an example, a parcellation of the plane (e.g. 2D grid complex) is always orientable.
The same is true for a tetrahedron, where the triangles divide the inner and outer side.
A Möbius strip (including a \CC{} discretization) is not orientable.
A cell complex where three triangles have one shared edge is also not orientable as it has three sides that are pairwise opposite of each other.

\begin{figure*}[t]
  \fontsize{9pt}{11pt}\selectfont
  \centering
  \def\svgwidth{\textwidth}
  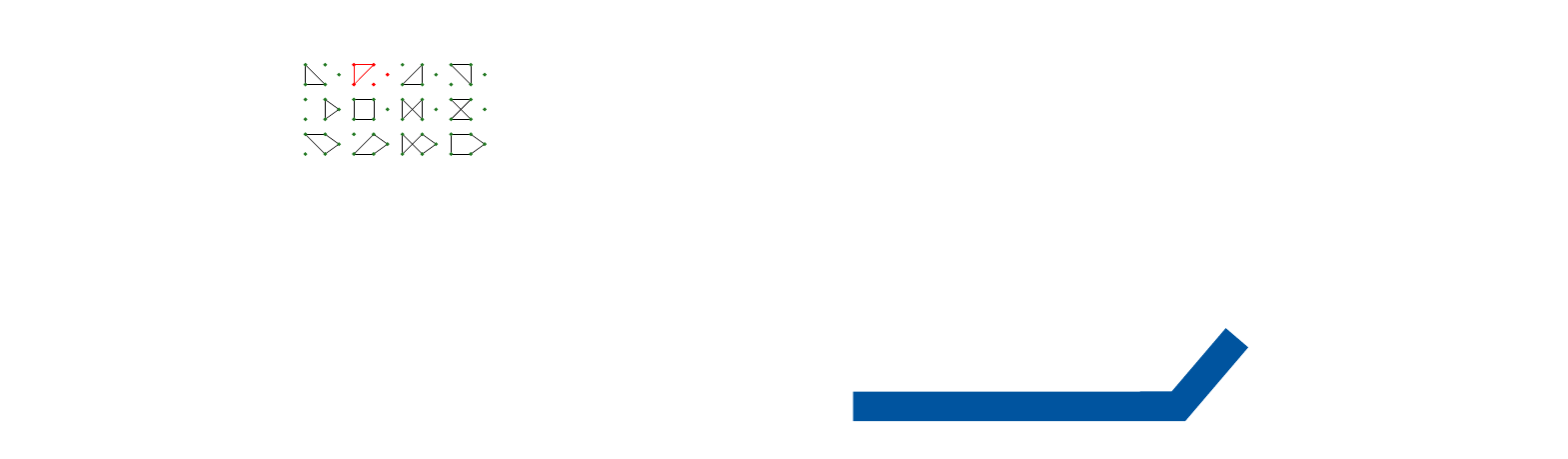
  \caption{Random lifting from 1-dim.~\CC{} to 2-dim.~\CC{} and our sampling algorithm. The model (strong background colors) is simulated by dividing it into two steps (light background colors): First, $s$ uniform spanning trees are sampled on the graph, each inducing a subset of all cycles. Depending on the probability $\rho_c$ for any cycle to appear in such a subset, the cycles are then sampled as cells. The algorithm (boxes) closely follows the two-step sampling model, approximating $\rho_c$ for efficiency.
  }
  \label{fig:sampling}
\end{figure*}

\subsection{Spanning-Tree-Based Cycle Sampling} 
In \cite{hoppe2026cycles}, we introduced a novel algorithm that samples spanning trees and simple cycles induced by the non-tree edges.
For each cycle $c$, there is a probability $\rho_c$ that $c$ is included in the set of cycles $C_T$ that is induced by a uniform spanning tree $T$.
Importantly, we provide an approximation for the probability of any cycle being induced that can be efficiently calculated over the spanning tree.
For the purposes of this paper, we will think of this algorithm as a black box that provides an estimator for the number of cycles (separately for each length), and that samples a subset of all cycles with skewed and non-independent but known inclusion probabilities $\rho_c$.
On most graph classes, cycles that are longer than $\Theta(\sqrt{n})$ are exponentially difficult to find even if they exist, and we carry this limitation over to our sampling algorithm for random cell complexes.
For more details on this, we refer the reader to \cite{hoppe2026cycles} rather than repeating the results here.

\section{A Model for Random Abstract Cell Complexes} \label{sec:model}

Random abstract cell complexes can be sampled through a series of iterative liftings that each lift a $k$-dimensional complex to a $(k+1)$-dimensional complex, as illustrated in \cref{fig:model}.
The first lifting, from a $0$-dimensional cell complex (set) to a $1$-dimensional cell complex (graph), adds $1$-cells (edges) with a fixed probability and is equivalent to the \erdosrenyi{} model.
For two- and higher-dimensional cells, the number of possible cells depends on the size of the boundary (e.g., edges for 2-cells).
Due to the combinatorial growth, possible 2-cells (simple cycles) are concentrated around long lengths in most graphs that are not too sparse.
To avoid only sampling cells with very large boundaries, for every dimension $k$ and possible boundary size $l$, we define the inclusion probability $P^{(k)}_l$.
We define our random cell complex model \RACC{}$(n, P^{(1)}, \dots, P^{(d_{\max})})$, where the vector $P^{(k)} = (P^{(k)}_l)$ collects the inclusion probabilities of $k$-cells with boundary size $l$ (cf.~\cref{alg:racc}).
Each possible cell is added independently according to the inclusion probability (i.e., depending on the boundary size).
For dimension $k=2$ and above, the cells that can be added depend on the sampled cells from dimension $k-1$.
This iterative lifting-based model also allows us to augment any given \CC{} with random higher-dimensional cells of specified dimensions.

\paragraph*{Relationship to the \erdosrenyi{} model} For 1-cells (edges), the only possible boundary size is $2$.
Therefore, the model \RACC{}$(n, P^{(1)})$ is equal to $G(n, P_2^{(1)})$.

\paragraph*{Relationship to the Linial--Meshulam and multi-parameter random \SC{} model} \RACC{} can be seen as a generalization of the \MRSC{} \cite{costa2016random} model, which itself generalizes the \LM{} \cite{linial2006homological} model.
The \LM{} model can be recovered using \RACC{} (non-mentioned inclusion probabilities are $0$):
\LM{} adds all possible simplices up to dimension $d_{\max}-1$ ($P^{(k)}_{k+1} = 1$ for $k < d_{\max}$), and adds $d_{\max}$-dimensional simplices with probability $p$ ($P^{(d_{\max})}_{d_{\max}+1} = p$).
\MRSC{} is equivalent to setting $P^{(k)}_{k+1} \leq 1$ instead.

\paragraph*{Sampling $2$-cells on a graph} As a step toward a sampling algorithm for the full model described above, we consider algorithms for the lifting from one-dimensional \CC{}s (graphs) to two-dimensional \CC{}s.
As there is no risk of confusion in this context, we simply use $P_l$ to refer to $P^{(2)}_l$ in the following.

Before we introduce our sampling algorithm, we discuss different ways to configure the model for practical applications.

\subsection[Heterogeneity for Static Target Probability P\_l]{Heterogeneity for Static Target Probability \texorpdfstring{$P_l$}{P\_l}}
\label{sec:alg:cycle-count}

Different realizations of \erdosrenyi{} graphs with the same parameters can differ a lot in the number of cycles for each length.
Thus, if we sample $2$-cells with a fixed probability $P_l$, there will be a similar variance in the number of sampled $2$-cells.
This behavior leads to variations in the dimension of the curl space, connectivity, and eigenvalues of the Hodge Laplacian $L_1$.
If required, this behavior can be changed by adjusting $P^{(2)}$ based on the realization of the 1-skeleton.
For this, we can estimate the number of cycles using the aforementioned spanning-tree-based algorithm \cite{hoppe2026cycles}.
Our package \texttt{py-raccoon} implements this already.

\subsection{RACC with an Exact Number of Cells}

Similar to the $G(n,m)$ version of \erdosrenyi, it may be useful to sample an exact number of $2$-cells rather than keeping to an expected number.
This can be achieved by sampling at least the desired number uniformly and then uniformly subsampling from the resulting set (for each length).
\texttt{py-raccoon}'s implementation also supports this directly, and our experiments use it as well.

\subsection{RACC with Non-Uniform Inclusion Probabilities}

So far, we have only discussed a uniform probability of $2$-cells (based on length), akin to the \erdosrenyi{} model for graphs.
However, we can derive biased models that prefer some $2$-cells over others.
For this, we would generalize from a length-based target inclusion probability $P_l$ to an individualized\footnote{or differently classified} inclusion probability $P_c$ for every cell $c$.
Analogously to the Stochastic Block Model, one could define multiplicative weights over the edges or nodes that modify the intended sampling probability of each potential cell.
For example, given node weights $w_u$ could then set $P_c \propto \prod_{u \in c} w_u$.
In this example, a straightforward calculation on spanning trees is possible, similar to calculating the length \cite{hoppe2026cycles}.
For any bias with this property, biased sampling will not change the computational cost of spanning-tree-based sampling.

However, we do not see a unique clear path to generalizing existing graph models.
For example, the stochastic block model could be lifted to $2$-cells based on either edge properties or node properties, and the properties could be multiplicative, additive, or aggregated in another way to obtain the desired probability.
To be useful in practice, the sampling probability has to be defined depending on the statistics to fix (exactly or in expectation).
Due to the large space of possible models, \texttt{py-raccoon} does not implement this directly.
However, it exposes an API for sampling spanning trees that can be used to easily implement the desired behavior.

\subsection{A simple rejection sampler for \RACC{}}
Before we discuss our more sophisticated sampler, we consider using rejection sampling.
First, we sample $l$ nodes at random (without replacement).
Then, we check if the resulting tuple is a valid cycle on the graph.
If yes, we accept, otherwise, we reject.
We repeat this process until we have sampled the desired number of $2$-cells or until number of samples $s$ relative to the number of possible permutations corresponds to the target probability $P_l$:
\[
P_l = \frac{s}{\prod_{i=0}^{l-1} n - i}
\]
This sampler is efficient for short cycles on small, dense graphs.
In other cases, it requires exponentially many samples.

\section[An Algorithm for Sampling 2-cells on a Graph]{An Algorithm for Sampling \texorpdfstring{$2$}{2}-Cells on a Graph}\label{sec:cell-sampling-alg}

We now introduce the sampling algorithm (\cref{alg:racc}) that, given an overall probability $P_l$ to sample a $2$-cell of length $l$, returns $2$-dimensional cell complex $\mathcal C$ s.t. $\forall c \in C_l: \mathbb P(c \in \mathcal C) = P_{|c|}$.
As we discuss below, the inclusion probabilities are independent if the number of samples is sufficiently large.
The algorithm can be used with either the exact inclusion probability $\rho_c$ or the mean field approximation $\rho_c^{\mathrm{MF}}$, with the latter introducing a sampling bias depending on the graph (see \cite{hoppe2026cycles}).

For this, we sample $s$ spanning trees as described above.
The algorithm from \cite{hoppe2026cycles} samples a spanning tree $T$, where each node $u$ has a depth (or distance from the root).
Each non-tree edge $(u,v)$ induces a cycle $c$, and the algorithm provides a list of tuples $(u, v, \lca(u,v), \rho_c)$  describing the cycle and listing the lowest common ancestor $\lca(u,v)$ in $T$.
From this, we calculate $|c|= \depth(u) + \depth(v) - 2\cdot\depth(\lca(u,v)) + 1$.
Then, the algorithm selects each induced cycle with probability $q_c$ s.t. the overall probability is $P_l$.
Each cycle $c$ has an independent probability of $\rho_cq_c$ to be selected from each of the $s$ spanning trees (and is only returned once if sampled multiple times).
Hence, we obtain
\begin{align}
  P_l = & 1 - (1-\rho_cq_c)^s \approx \rho_cq_cs\\
\Leftrightarrow\quad q_c =& \frac{1-\sqrt[s]{1-P_l}}{\rho_c} \approx \frac{P_l}{\rho_cs}.
\end{align}

This equation only has a valid solution for $1-\sqrt[s]{1-P_l} \leq \rho_c$, i.e., only for sufficiently small target probabilities $P_l$ or sufficiently large $s$.
Otherwise, we will accept all found cycles, but still find fewer cycles than we intended to sample.
In other words, for either very large numbers of sampled $2$-cells or for lengths which appear very infrequently, we have to increase the number of sampled spanning trees.
For cycles of order $\mathcal O(\sqrt{n})$, this cost is small as they occur frequently in uniform spanning trees.
Longer cycles are induced with decreasing frequency, leading to an exponential number of samples required.

This is a direct result of the limited height of the tree, which is $\Theta(\sqrt{n})$ in expectation for many common graph classes \cite{hoppe2026cycles}.
In turn, this is an expression of the overall NP-hardness of finding cycles of arbitrary length.

\begin{algorithm}[t]
    \caption{\RACC{} iterative lifting algorithm.}\label{alg:racc}
    \textbf{Input:} \CC{} $\mathcal C, d_{\max{}}, P^{(k)} \text{ for } \dim \mathcal C < k \leq d_{\max{}}$
    \begin{algorithmic}
    \For{$k \gets (\dim \mathcal C + 1)..d_{\max{}}$}
        \For{$l$ s.t. $P^{(k)}_l > 0$}
            \State $X \gets \forall$ possible cells $c$ with boundary size $l:$ select independently with probability $P_l^{(k)}$
            \State $\mathcal C \gets \mathcal C \cup X$
        \EndFor
    \EndFor \\
    \Return $\mathcal C$
    \end{algorithmic}
\end{algorithm}

\subsection{Independence and Sample Size}
The inclusion of different cells is independent for sufficiently many sampled trees $s$.
Otherwise, multiple cells may be sampled from the same spanning tree.
Generally, different cycles in a spanning tree are not independent.
Cycles that are long relative to the height of the tree also have a relevant chance of sharing an edge.
Short cycles, on the other hand, are more locally contained and less likely to share an edge.
To ensure independence of sampled $2$-cells, the number of sampled trees $s$ has to be sufficiently large.

\Cref{fig:independence} showcases how this affects the biggest eigenvalue $\lambda_{\max}$ of the up Laplacian $L_1^{\text{up}}$.
Where the number of sampled trees is smaller than the number of sampled cells, the shared spanning tree edges lead to a higher overlap and thus bigger eigenvalues.
For $s = N_5 = 100$, we can see that the behavior is already changed substantially.
For bigger $s$, there is much less change than below that threshold, indicating that using as many trees as there are desired cells results in an existing, but small correlation of $2$-cells.

\subsection{Effect of the Mean Field Approximation}

\begin{figure}[t]
    \centering
    \includegraphics[width=\linewidth]{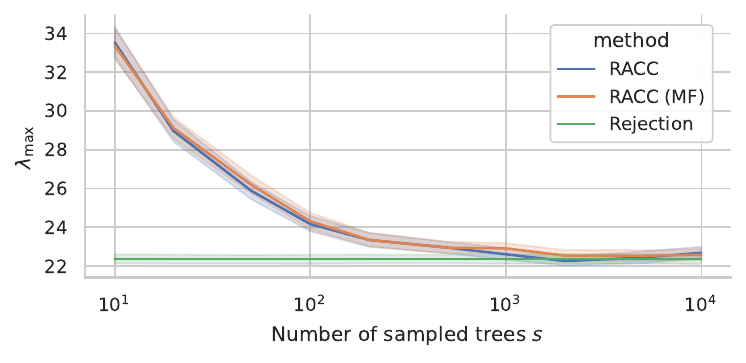}
    \caption{Largest eigenvalue $\lambda_{\max{}}$ (mean, 95\%-confidence interval) of the up Laplacian $L_1^{\text{up}}=B_2 B_2^\top$ of 100 random cell complexes per configuration with $N_5 = 100$ $2$-cells of length $5$ on a shared $G(20, 0.5)$ skeleton, for different numbers of sampled trees $s$. We compare \RACC{} with exact inclusion probabilities $\rho_c$ (\RACC{}) and with the mean field approximation $\rho_c^{\mathrm{MF}}$ (\RACC{} (\textsc{mf})) to cells obtained using rejection sampling (Rejection), which does not depend on $s$.}\label{fig:independence}
\end{figure}

The mean-field approximation \cite{hoppe2026cycles} assumes that all nodes are equally connected.
In some graph classes (e.g. 2D grid graphs) with large diameters and locally concentrated connectivity, there is a large bias that applies to all cycles of the same length.
There is also an error component in the probability approximation that is different between cycles of the same length.
The latter was far less pronounced in the graphs that were considered \cite{hoppe2026cycles}.

The length-based bias has a strong effect when sampling using the direct inclusion probability $P_l$.
On the other hand, when sampling an exact number of $2$-cells, this bias also appears in the cycle count estimate, canceling out overall.

In our example in \cref{fig:independence}, we sampled exact numbers of 2-cells, meaning only the smaller, relative bias exists.
Consistent with the experiments in \cite{hoppe2026cycles}, this does not result in a significant difference in the eigenvalue compared to simple rejection sampling, which does not have these issues.

\begin{figure*}[p]
    \begin{minipage}{.48\linewidth}
        \centering
        \includegraphics[width=.8\linewidth]{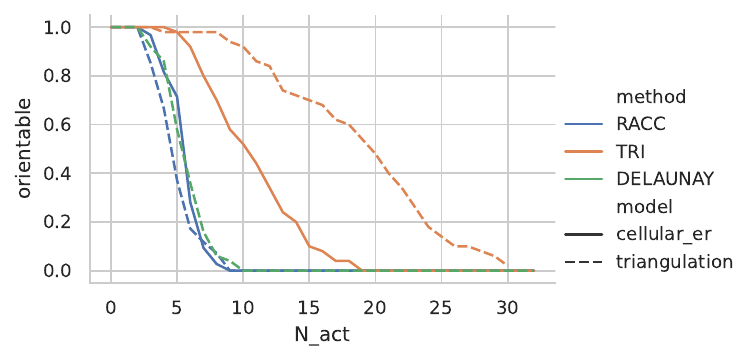}
        \panellabel{a}
    \end{minipage}
    \begin{minipage}{.48\linewidth}
        \centering
        \includegraphics[width=\linewidth]{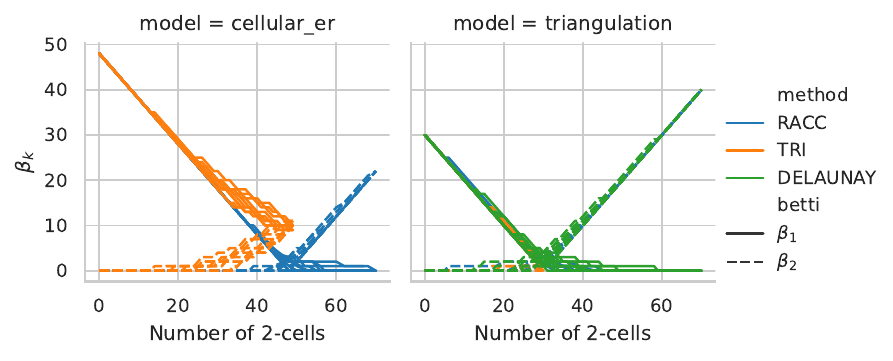}
        \panellabel{b}
    \end{minipage}\hfill
    \caption{Analysis of random cell complexes, comparing our model (\RACC{}; exact $\rho_c$, $s=200$ trees) to the triangulation-based model from \cite{hoppe2023representing} (\textsc{delaunay}) and randomly selected triangles (\textsc{tri}); 50 runs each. Comparison on both a fixed graph sampled from $G(n,p)$ ($n=20, p=0.3$, $67$ edges) and a fixed Delaunay triangulation ($n=20$, $49$ edges).
    (a) Fraction of sampled \CC{}s that are orientable. (b) Comparison of the 1- and 2-cohomology of cell complexes; 10 runs shown, one line per run.
    }
    \label{fig:props}
\end{figure*}

\begin{figure*}[p]
    \begin{minipage}{.49\linewidth}
        \centering
        \includegraphics[width=\linewidth]{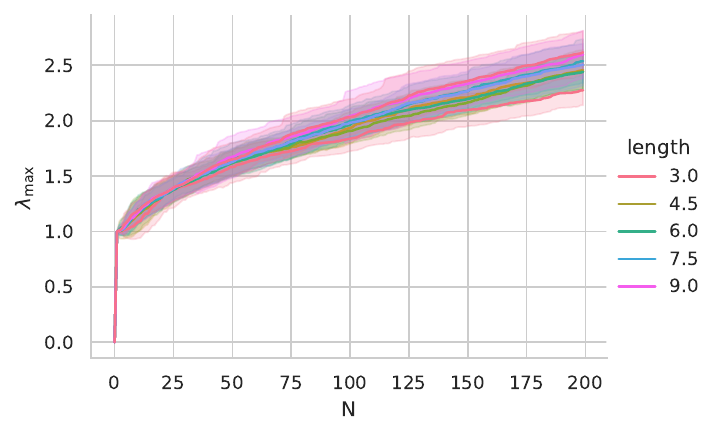}
        \panellabel{a}
    \end{minipage}
    \hfill
    \begin{minipage}{.49\linewidth}
        \centering
        \includegraphics[width=\linewidth]{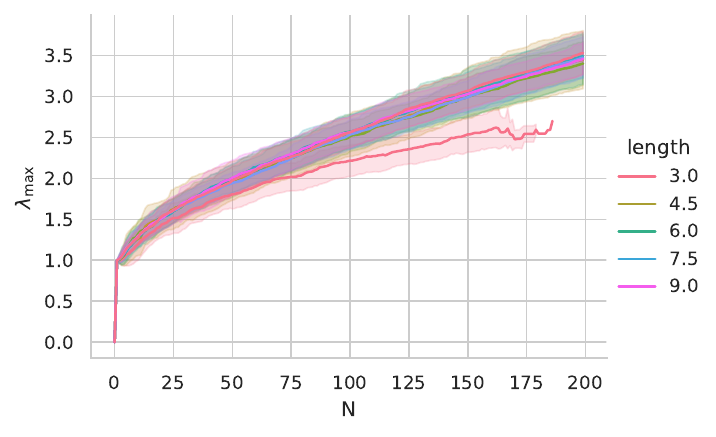}
        \panellabel{b}
    \end{minipage}
    \caption{Largest eigenvalues $\lambda_{\max}$ of the normalized up Laplacians $\mathcal L_1^{\text{up}}=B_2 D_2^{-1} B_2^\top$ of sampled \CC{}s on a (a) dense $G(100, 0.2)$ and (b) sparse $G(100,0.1)$ skeleton. In both panels, we compare \CC{}s sampled from \emph{\RACC{}} (exact $\rho_c$, $s=400$ trees) configured to only sample cells of length $l$. The degree matrix $D_2 = \mathrm{diag}(|B_2|^\top\mathbf{1})$ contains the number of elements in the boundary of each $2$-cell.}\label{fig:lapl_eig}
\end{figure*}

\begin{figure*}[p]
    \centering
    \includegraphics[width=.85\linewidth]{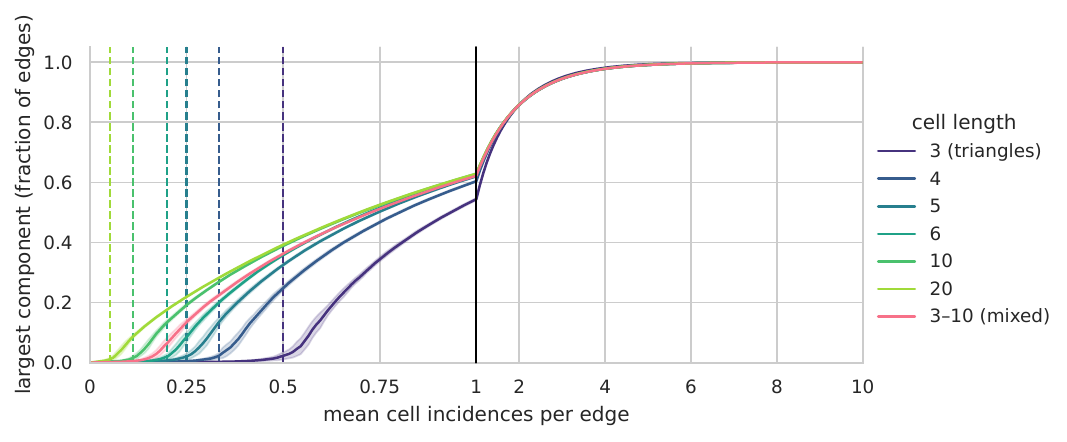}
    \caption{Higher-order percolation (relative size of the largest component) with different cell boundary sizes $l$, mean and standard deviation over 50 runs on different $G(n,m)$-graphs with $n=500, m=20~000$. \RACC{} with mean field $\rho_c^{\mathrm{MF}}$, $s=60~000$ trees; mixed lengths $3$--$10$ in equal proportion. Vertical dashed lines mark the theoretical percolation thresholds $\nicefrac{1}{l-1}$.}\label{fig:percolation}
\end{figure*}

\section{Properties of Random Cell Complexes}\label{sec:model-properties}

Before using \RACC{} in applications, it is important to understand its basic properties.
In this section, we explore the properties of the model, and, where applicable, compare to simplicial complexes that are generated from the same underlying graph.

First, in \cref{fig:props}(a), we can see that most obtained \CC{}s are not orientable, even with relatively few added cells.
This is in contrast to \SC{}s, which are orientable with considerably more added triangles, especially on triangulation graphs.
Second, \cref{fig:props}(b) shows that in our experiments random \CC{}s either have (almost) no 1-cohomology or (almost) no 2-cohomology.
Simplicial complexes obtained on the same graphs often do not remove all 1-cohomologies; and are more limited in the number of 2-cohomologies they add.
Since this depends on the density of triangles in the graph, this effect is especially pronounced on sparse random graphs and less or not present on dense graphs.
Our model behaves similarly on graphs obtained from point cloud triangulations.
Due to the local structure, \SC{}s can fill all 1-cohomologies, but add only few 2-cohomologies; they also remain orientable for many added triangles.

The spectrum of the Hodge Laplacian reflects the connectivity and structure of a cell complex.
In \Cref{fig:lapl_eig}, we compare the largest eigenvalues of the normalized up Laplacian $\mathcal L_1^{\text{up}}$ of the null model and the closest equivalent simplicial complex, i.e., uniformly adding triangles.
We observe that, on a dense graph well-covered by triangles, all cell complexes (including the simplicial complexes) exhibit similar behavior.
However, on the sparser graph, there are not enough triangles, and the random simplicial complex has a significantly smaller eigenvalue compared to random cell complexes with longer $2$-cells.
This is due to the increased overlap between $2$-cells, enabling a closely connected complex overall.

In other words, triangles and other $2$-cells capture the connectivity of the graph similarly, as long as the graph has sufficient triangles.
On sparser graphs, (random) cell complexes uniquely enable the same behavior.
In this case, we posit that it is relevant to cover the entire cycle space (which spanning-tree-based methods always do), but not necessarily to cover cycles of arbitrary length (which no method can accomplish efficiently).

Finally, we look at percolation through higher-order paths \cite{dhara2026giants}.
In this definition, two edges are adjacent if they have a shared coface ($2$-cell).
Percolation is characterized by the upper incidence (mean number of cofaces) of edges $\bflambda = \mathbf{1}^\top |B_2| \mathbf{1} \cdot |\mathcal C_1|^{-1}$ (where $|B_2|$ is the element-wise absolute value of $B_2$) \cite{dhara2026giants}.

The authors showed that for $\bflambda > \nicefrac{1}{2}$, a giant component starts forming.
For triangles, every edge is adjacent to two neighbors.
With the same branching argument \cite{dhara2026giants}, it follows that the threshold is always at $\bflambda > \nicefrac{1}{l-1}$, where $l$ is the size of the boundary of the $2$-cell.
Triangles are $2$-cells with $l=3$, thus the special case of triangles holds with $\nicefrac{1}{l-1} = \nicefrac{1}{2}$.
With more neighbors per upper-incident cell, according to the hypothesis, we require fewer cells to achieve percolation.
Equivalently, if we re-formulate this as the degree $d_\mathrm{up}$ (number of neighbors) of the higher-order connection, we arrive at a threshold of $d_{\mathrm{up}} > 1$, which reflects percolation on the \erdosrenyi{} random graphs.
\Cref{fig:percolation} shows empirical percolation on a number of large random graphs.
The hypothetical percolation thresholds are consistent with the empirical behavior we observe.
Beyond the percolation threshold, the sizes of the largest connected component obtained from different cell sizes converge.
One noteworthy difference not visible in the plot is that for sparse random graph configurations, there are not enough triangles to connect the majority of edges (or even achieve percolation).

Both the cohomology and spectral analysis show a key shortcoming of approaches limited to triangles:
On many graphs, triangles do not span the entire graph as there are few and often concentrated in smaller cliques.
Given that these properties are highly dependent on the underlying graph, these analyses only provide a first intuitive insight into random \CC{}s.
Thus, further theoretical analysis is required to fully understand the properties of random \CC{}s.

\section{Applications} \label{sec:applications}

In this section, we explore the use cases for random cell complexes we identified earlier.
This demonstrates that both the model and the sampling algorithm are fit to close the research gap we identified earlier.

\begin{figure}[t]
    \centering
    \includegraphics[width=\linewidth]{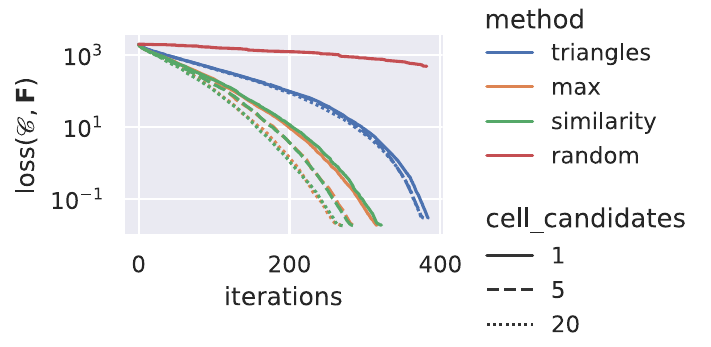}
    \caption{Comparison of the approaches to the flow representation learning problem from \cite{hoppe2023representing} and random $2$-cells (random) on the Anaheim dataset \cite{gh-transportation-networks}. The random $2$-cells are sampled using \RACC{} with exact inclusion probabilities $\rho_c$ ($s=2000$ sampled trees), with lengths $3$ to $6$ following the length distribution of the cells inferred by \cite{hoppe2023representing}.
    }
    \label{fig:baseline-flow-representation}
\end{figure}

\subsection{Null Model as Experiment Baseline}

When evaluating novel methods that infer cells, it is useful to have a baseline that adds cells without using any information (maximising randomness).
If a method infers meaningful structure, we expect it to significantly outperform the random baseline.
In the case of the flow representation learning problem \cite{hoppe2023representing}, we see that random 2-cells are very inefficient in \cref{fig:baseline-flow-representation}.
However, this also gives us a way to classify the performance of the approaches we want to evaluate.
In this case, we see that the proposed methods from \cite{hoppe2023representing} give a representation that provides a significant improvement over \textit{triangles}, but the triangle baseline already explains much of the structure (compared to random $2$-cells).

\subsection{Sensitivity Analysis: How Important are Higher-Order Interactions?}

In this section, we demonstrate the use of random \CC{}s for the sensitivity analysis (or ablation analysis~\cite{fawcett2016analysing}) of neural networks defined on higher-order networks.

In \cite{bodnar2021weisfeiler}, Bodnar et al.\ proposed an architecture for cell complex neural networks (`\CWN{}s').
To obtain a cell complex and capture structural information on a graph, they add chordless cycles as 2-cells.
Here we use our random cell complex model to investigate to what extent this specific choice of 2-cells is relevant or if adding random cells would have yielded similar results.
This provides insights into whether it is the \emph{specific} cell structure (adding chordless cycles) that confers an advantage, or whether it is merely the increase in expressive power of the neural networks (resulting from adding higher-order cells) that is of importance.

\begin{table}[t]
\caption{Comparison of \CWN{} \cite{bodnar2021weisfeiler} with the original 2-cells (`Rings'), random 2-cells (`\RND{}'), and no 2-cells (`None'). For `\RND{}', we sample the same number of 2-cells of each length as `Rings' (lengths up to 6) using \RACC{} with exact inclusion probabilities $\rho_c$. Code available at \cwnrepolink{}. Results deviate slightly from \cite{bodnar2021weisfeiler} despite using the original code. Accuracy of \PPGN{}s taken from \cite{bodnar2021weisfeiler}.} 
    \label{tbl:cwn}
    \centering
    \begin{tabular}{lcc}
        \toprule
        Dataset \cite{morris2020tudataset} & \PROTEINS{} & \NCI{} \\
        \midrule
        \CWN{} (Rings) & $72.7 \pm 4.3$ & $\mathbf{84.4 \pm 1.5}$\\
        \CWN{} (\RND{}) & $74.1 \pm 5.0$ & $\mathbf{84.6 \pm 1.5}$ \\
        \CWN{} (None) & $73.0 \pm 5.5 $ & $82.9 \pm 1.2$ \\
        \midrule
        \PPGN{}s \cite{maron2019provably} & $\mathbf{77.2\pm 4.7}$ & $82.2 \pm 1.4$\\
        \bottomrule
    \end{tabular}
\end{table}

Interestingly, for the \PROTEINS{} dataset (see \cref{tbl:cwn}) there is no significant performance difference if we simply add random 2-cells instead of following the procedure outlined in~\cite{bodnar2021weisfeiler}. 
Indeed, a similar performance can even be achieved by not adding any 2-cells at all.
On the \NCI{} dataset \cite{morris2020tudataset}, we also observe that random $2$-cells lead to the same result as the original lifting.
However, a closer inspection shows that for the lengths considered in the evaluation code (up to 6), all simple cycles are already chordless, i.e., there are no degrees of freedom for the null model and we obtain the same lifting.\footnote{All datasets where \CWN{} performs best among the tested approaches have a similarly small number of edges compared to nodes, likely resulting in a similar behavior. Performing further experiments that were not part of the original paper would be interesting, but out of scope for this work.}
On both graphs, we conclude that the 2-cell lifting procedure proposed in~\cite{bodnar2021weisfeiler} is not specifically required for the observed performance as it fails to capture relevant topological information on \PROTEINS{} beyond what we could expect by simply increasing the expressivity of the network.
To corroborate our findings, we have added a further comparison to the \PPGN{} architecture~\cite{maron2019provably}.
\PPGN{} also features an increased expressivity compared to standard \GNN{}s, but is not based on any topological considerations.
As~\cref{tbl:cwn} shows \PPGN{} works well in the \PROTEINS{} dataset, yet only performs as well as  \CWN{} without $2$-cells on \NCI{}.
What additional information can we take away from these findings?
The direct findings can motivate interesting future analysis.
On \PROTEINS{}, there may be a different cell complex lifting that leads to better results and matches \PPGN{}s; and on \NCI{}, considering longer cycles may yield better performance and more insights into the lifting procedure.

\section{Conclusion}

The main contributions of this paper are (i) the introduction of a simple model for random abstract cell complexes, and (ii) an efficient approximate sampling algorithm for the model.
We explored the orientability, eigenvalues, and percolation on random cell complexes and showcased how the model enables a number of investigations concerning higher-order network data and learning methods operating with such data:
Specifically, random \CC{}s can be used as synthetic input data and as null models for multiple purposes.
In contrast to previous maximum-entroppy models for cell complexes, \RACC{} is both unbiased and able to lift arbitrary $1$-skeletons to $2$-dimensional cell complexes.

This groundwork opens up multiple avenues for future research on abstract cell complexes:
First, a more theoretical analysis of the number of simple cycles and the behavior of \RACC{}s.
Second, while we defined our model for arbitrary dimensions, higher-dimensional cells add considerable combinatorial complexity, which is why we limited our sampling algorithm to 2-cells.
Indeed, adapting our sampling algorithm to higher-order cells is not trivial as the higher-dimensional analogs of spanning trees do not share many of their desirable properties.

When employing random cell complexes as synthetic data, it may also be desirable to include more structure in the cell complexes, both on the edge and cell levels.
On the edge level, our lifting already allows for arbitrary graphs (including specific random models).
For the cell level, we sketched how this could be achieved, but the properties to target in a model have to be adjusted for the specific application.
Such adaptations make the model suited for a variety of uses.

\section*{Code and Data Availability}
The implementation of the sampling algorithm and the evaluation code are available at \repolink{}.
The sampling algorithm is also distributed as the \href{https://pypi.org/project/py-raccoon/}{\texttt{py-raccoon}} package on the Python Package Index.
The code for the \CWN{} experiments is available at \cwnrepolink{}.
The Anaheim network is taken from the Transportation Networks for Research repository~\cite{gh-transportation-networks}; \PROTEINS{} and \NCI{} are taken from TUDataset~\cite{morris2020tudataset}.

\section*{AI usage declaration}
AI was used to assist with writing code for data visualization and handling of computational experiments.
All algorithmic functionality was implemented by hand, and all AI-generated code was carefully vetted.
The authors also used generative AI to assist with grammatical and language editing of the manuscript as well as revision of passages; and to find potential errors in the arguments and exposition.
All outputs and edits were critically reviewed, manually verified, and revised by the authors.
The authors claim full responsibility for the final version of the manuscript.

\begin{acknowledgments}
Funded by the European Union (ERC, HIGH-HOPeS, 101039827). Views and opinions expressed are however those of the authors only and do not necessarily reflect those of the European Union or the European Research Council Executive Agency. Neither the European Union nor the granting authority can be held responsible for them.
\end{acknowledgments}

\bibliography{main}

@article{fosdick2018configuring,
  title     = {Configuring random graph models with fixed degree sequences},
  author    = {Fosdick, Bailey K and Larremore, Daniel B and Nishimura, Joel and Ugander, Johan},
  journal   = {Siam Review},
  volume    = {60},
  number    = {2},
  pages     = {315--355},
  year      = {2018},
  publisher = {SIAM}
}

@article{battiston2020networks,
  title     = {Networks beyond pairwise interactions: Structure and dynamics},
  author    = {Battiston, Federico and Cencetti, Giulia and Iacopini, Iacopo and Latora, Vito and Lucas, Maxime and Patania, Alice and Young, Jean-Gabriel and Petri, Giovanni},
  journal   = {Physics Reports},
  volume    = {874},
  pages     = {1--92},
  year      = {2020},
  publisher = {Elsevier}
}

@article{bick2023higher,
  title     = {What are higher-order networks?},
  author    = {Bick, Christian and Gross, Elizabeth and Harrington, Heather A and Schaub, Michael T},
  journal   = {SIAM Review},
  volume    = {65},
  number    = {3},
  pages     = {686--731},
  year      = {2023},
  publisher = {SIAM}
}

@article{torres2021and,
  title     = {The why, how, and when of representations for complex systems},
  author    = {Torres, Leo and Blevins, Ann S and Bassett, Danielle and Eliassi-Rad, Tina},
  journal   = {SIAM Review},
  volume    = {63},
  number    = {3},
  pages     = {435--485},
  year      = {2021},
  publisher = {SIAM}
}

@article{wasserman2018topological,
  title     = {Topological data analysis},
  author    = {Wasserman, Larry},
  journal   = {Annual Review of Statistics and Its Application},
  volume    = {5},
  pages     = {501--532},
  year      = {2018},
  publisher = {Annual Reviews}
}

@article{barbarossa2020topological,
  title     = {Topological signal processing over simplicial complexes},
  author    = {Barbarossa, Sergio and Sardellitti, Stefania},
  journal   = {IEEE Transactions on Signal Processing},
  volume    = {68},
  pages     = {2992--3007},
  year      = {2020},
  publisher = {IEEE}
}

@article{schaub2021signal,
  title     = {Signal processing on higher-order networks: Livin’on the edge... and beyond},
  author    = {Schaub, Michael T and Zhu, Yu and Seby, Jean-Baptiste and Roddenberry, T Mitchell and Segarra, Santiago},
  journal   = {Signal Processing},
  volume    = {187},
  pages     = {108149},
  year      = {2021},
  publisher = {Elsevier}
}

@article{hajij2022topological,
  title   = {Topological deep learning: Going beyond graph data},
  author  = {Hajij, Mustafa and Zamzmi, Ghada and Papamarkou, Theodore and Miolane, Nina and Guzm{\'a}n-S{\'a}enz, Aldo and Ramamurthy, Karthikeyan Natesan and Birdal, Tolga and Dey, Tamal K and Mukherjee, Soham and Samaga, Shreyas N and others},
  journal = {arXiv preprint arXiv:2206.00606},
  year    = {2022}
}

@article{drobyshevskiy2019random,
  title     = {Random graph modeling: A survey of the concepts},
  author    = {Drobyshevskiy, Mikhail and Turdakov, Denis},
  journal   = {ACM computing surveys (CSUR)},
  volume    = {52},
  number    = {6},
  pages     = {1--36},
  year      = {2019},
  publisher = {ACM New York, NY, USA}
}

@article{albert2002statistical,
  title     = {Statistical mechanics of complex networks},
  author    = {Albert, R{\'e}ka and Barab{\'a}si, Albert-L{\'a}szl{\'o}},
  journal   = {Reviews of modern physics},
  volume    = {74},
  number    = {1},
  pages     = {47},
  year      = {2002},
  publisher = {APS}
}

@incollection{costa2016random,
  title     = {Random simplicial complexes},
  author    = {Costa, Armindo and Farber, Michael},
  booktitle = {Configuration Spaces: Geometry, Topology and Representation Theory},
  pages     = {129--153},
  year      = {2016},
  publisher = {Springer}
}

@inproceedings{hoppe2023representing,
  title     = {{Representing Edge Flows on Graphs via Sparse Cell Complexes}},
  author    = {Josef Hoppe and Michael T. Schaub},
  booktitle = {The Second Learning on Graphs Conference},
  year      = {2023},
  url       = {https://openreview.net/forum?id=qix189lq5D}
}

@article{giscard2017evaluating,
  title     = {Evaluating balance on social networks from their simple cycles},
  author    = {Giscard, Pierre-Louis and Rochet, Paul and Wilson, Richard C},
  journal   = {Journal of Complex Networks},
  volume    = {5},
  number    = {5},
  pages     = {750--775},
  year      = {2017},
  publisher = {Oxford University Press}
}

@misc{gh-transportation-networks,
  author = {{Transportation Networks for Research Core Team}},
  url    = {https://github.com/bstabler/TransportationNetworks},
  note   = {\href{https://github.com/bstabler/TransportationNetworks}{\texttt{github.com/\allowbreak{}bstabler/\allowbreak{}TransportationNetworks}}. Accessed: 2024-01-29.},
  title  = {Transportation Networks for Research},
  year   = {2020}
}

@article{bodnar2021weisfeiler,
  title   = {{Weisfeiler and Lehman Go Cellular: CW Networks}},
  author  = {Bodnar, Cristian and Frasca, Fabrizio and Otter, Nina and Wang, Yuguang and Lio, Pietro and Montufar, Guido F and Bronstein, Michael},
  journal = {Advances in Neural Information Processing Systems},
  volume  = {34},
  pages   = {2625--2640},
  year    = {2021}
}

@inproceedings{sardellitti2021topological,
  title        = {Topological signal processing over cell complexes},
  author       = {Sardellitti, Stefania and Barbarossa, Sergio and Testa, Lucia},
  booktitle    = {2021 55th Asilomar Conference on Signals, Systems, and Computers},
  pages        = {1558--1562},
  year         = {2021},
  organization = {IEEE}
}

@article{morris2020tudataset,
  title   = {{TUDataset}: A collection of benchmark datasets for learning with graphs},
  author  = {Morris, Christopher and Kriege, Nils M and Bause, Franka and Kersting, Kristian and Mutzel, Petra and Neumann, Marion},
  journal = {arXiv preprint arXiv:2007.08663},
  year    = {2020}
}

@misc{taxi-dataset,
  author = {Chris Whong},
  title  = {{FOILing NYC’s Taxi Trip Data}},
  note   = {\href{https://chriswhong.com/open-data/foil_nyc_taxi/}{\texttt{chriswhong.com/open-data/foil\_nyc\_taxi}}. Accessed: 2024-03-28.},
  year   = {2014}
}

@article{fawcett2016analysing,
  title     = {Analysing differences between algorithm configurations through ablation},
  author    = {Fawcett, Chris and Hoos, Holger H},
  journal   = {Journal of Heuristics},
  volume    = {22},
  pages     = {431--458},
  year      = {2016},
  publisher = {Springer}
}

@article{maron2019provably,
  title   = {Provably powerful graph networks},
  author  = {Maron, Haggai and Ben-Hamu, Heli and Serviansky, Hadar and Lipman, Yaron},
  journal = {Advances in neural information processing systems},
  volume  = {32},
  year    = {2019}
}

@article{erdos1959random,
  title   = {On random graphs {I}},
  author  = {Erd\H{o}s, Paul and R\'enyi, Alfr\'ed},
  journal = {Publ. math. debrecen},
  volume  = {6},
  number  = {290-297},
  pages   = {18},
  year    = {1959}
}

@article{gilbert1959random,
  title     = {Random graphs},
  author    = {Gilbert, Edgar N},
  journal   = {The Annals of Mathematical Statistics},
  volume    = {30},
  number    = {4},
  pages     = {1141--1144},
  year      = {1959},
  publisher = {JSTOR}
}

@article{holland1983stochastic,
  title     = {Stochastic blockmodels: First steps},
  author    = {Holland, Paul W and Laskey, Kathryn Blackmond and Leinhardt, Samuel},
  journal   = {Social networks},
  volume    = {5},
  number    = {2},
  pages     = {109--137},
  year      = {1983},
  publisher = {Elsevier}
}

@article{van2018sparse,
  title     = {Sparse maximum-entropy random graphs with a given power-law degree distribution},
  author    = {van der Hoorn, Pim and Lippner, Gabor and Krioukov, Dmitri},
  journal   = {Journal of Statistical Physics},
  volume    = {173},
  pages     = {806--844},
  year      = {2018},
  publisher = {Springer}
}

@article{linial2006homological,
  title     = {Homological connectivity of random 2-complexes},
  author    = {Linial*, Nathan and Meshulam*, Roy},
  journal   = {Combinatorica},
  volume    = {26},
  number    = {4},
  pages     = {475--487},
  year      = {2006},
  publisher = {Springer}
}

@article{zahle1988random,
  title     = {Random cell complexes and generalised sets},
  author    = {Zähle, Martina},
  journal   = {The Annals of Probability},
  pages     = {1742--1766},
  year      = {1988},
  publisher = {JSTOR}
}

@article{schweinhart2016topological,
  title     = {Topological similarity of random cell complexes and applications},
  author    = {Schweinhart, Benjamin and Mason, Jeremy K and MacPherson, Robert D},
  journal   = {Physical Review E},
  volume    = {93},
  number    = {6},
  pages     = {062111},
  year      = {2016},
  publisher = {APS}
}

@article{leistritz1992topological,
  title     = {Topological mean value relations for random cell complexes},
  author    = {Leistritz, Lutz and Z{\"a}hle, Martina},
  journal   = {Mathematische Nachrichten},
  volume    = {155},
  number    = {1},
  pages     = {57--72},
  year      = {1992},
  publisher = {Wiley Online Library}
}

@article{watts1998collective,
  title     = {Collective dynamics of ‘small-world’networks},
  author    = {Watts, Duncan J and Strogatz, Steven H},
  journal   = {nature},
  volume    = {393},
  number    = {6684},
  pages     = {440--442},
  year      = {1998},
  publisher = {Nature Publishing Group}
}

@article{dall2002random,
  title     = {Random geometric graphs},
  author    = {Dall, Jesper and Christensen, Michael},
  journal   = {Physical review E},
  volume    = {66},
  number    = {1},
  pages     = {016121},
  year      = {2002},
  publisher = {APS}
}

@article{kahle2014topology,
  title   = {Topology of random simplicial complexes: a survey},
  author  = {Kahle, Matthew},
  journal = {AMS Contemp. Math},
  volume  = {620},
  pages   = {201--222},
  year    = {2014}
}

@article{kahle2009topology,
  title     = {Topology of random clique complexes},
  author    = {Kahle, Matthew},
  journal   = {Discrete mathematics},
  volume    = {309},
  number    = {6},
  pages     = {1658--1671},
  year      = {2009},
  publisher = {Elsevier}
}

@misc{hoppe2026cycles,
  title         = {{Efficiently Estimating Cycle Statistics on Graphs via Spanning Trees}},
  author        = {Josef Hoppe and Michael T. Schaub},
  note          = {{arXiv Preprint. arXiv:2611.07925}},
  year          = {2026},
}

@misc{hoppe2025cellcomplexes,
  title         = {{Don't be Afraid of Cell Complexes! An Introduction from an Applied Perspective}},
  author        = {Josef Hoppe and Vincent P. Grande and Michael T. Schaub},
  note          = {{arXiv Preprint. arXiv:2506.09726}},
  year          = {2025},
  eprint        = {2506.09726},
  archiveprefix = {arXiv},
  primaryclass  = {eess.SP},
  url           = {https://arxiv.org/abs/2506.09726}
}

@article{dhara2026giants,
  title={Giants through higher-order paths in random simplicial complexes},
  author={Dhara, Souvik and Kang, Taegyu},
  journal={arXiv preprint arXiv:2605.03151},
  year={2026}
}

@article{alon2008biomolecular,
  title={Biomolecular network motif counting and discovery by color coding},
  author={Alon, Noga and Dao, Phuong and Hajirasouliha, Iman and Hormozdiari, Fereydoun and Sahinalp, S Cenk},
  journal={Bioinformatics},
  volume={24},
  number={13},
  pages={i241--i249},
  year={2008},
  publisher={Oxford University Press}
}

@article{valiant1979complexity,
  title={The Complexity of Enumeration and Reliability Problems},
  author={Valiant, Leslie G.},
  journal={SIAM Journal on Computing},
  volume={8},
  number={3},
  pages={410--421},
  year={1979},
  doi={10.1137/0208032}
}

@article{young2017construction,
  title={Construction of and efficient sampling from the simplicial configuration model},
  author={Young, Jean-Gabriel and Petri, Giovanni and Vaccarino, Francesco and Patania, Alice},
  journal={Physical Review E},
  volume={96},
  number={3},
  pages={032312},
  year={2017},
  publisher={APS}
}

@article{courtney2016generalized,
  title={Generalized network structures: The configuration model and the canonical ensemble of simplicial complexes},
  author={Courtney, Owen T and Bianconi, Ginestra},
  journal={Physical Review E},
  volume={93},
  number={6},
  pages={062311},
  year={2016},
  publisher={APS}
}
\bibliographystyle{apsrev4-2}

\clearpage
\appendix
\crefalias{section}{appendix}

\FloatBarrier
\section{Related Work}\label{app:relwork}

\subsection{Random Graph Models}
Random graph models started with the \erdosrenyi{} random graph model \cite{erdos1959random}, which originally proposed to draw a graph uniformly at random from the set of all graphs with a fixed number of nodes and edges.
In this paper, we refer to a variant introduced by Gilbert~\cite{gilbert1959random} but commonly also referred to as  \erdosrenyi{} model, that draws each edge independently with probability $p$.
More complex graph models generally focus on modeling community structure (most notably the Stochastic Block Model \cite{holland1983stochastic}) or the distribution of node degrees \cite{albert2002statistical,van2018sparse}.
Furthermore, there are models aimed at more closely approximating real-world networks, such as the Watts--Strogatz small-world graph \cite{watts1998collective}, modeling the small diameter of real-world networks.
Finally, graphs may be generated from random point clouds by connecting nodes that are close to each other in metric space \cite{dall2002random}; resulting in the formation of clusters similar to real-world networks.
There are many more models and ways to configure models that would exceed the scope of this literature review, but it should be noted that many use cases for these models exist, requiring different graph models.

\subsection{Random Simplicial Complex Models}
The Linial--Meshulam model for random \SC{}s \cite{linial2006homological} is an extension of \erdosrenyi{} to higher dimensions.
\LM{} takes a complete $(d-1)$-dimensional complex and samples i.i.d.~from all possible $d$-simplices.
The Linial--Meshulam model was generalized in multiple ways, e.g., a different static probability $P^{(k)}_{k+1}$ to sample $k$-simplices (separately) for each dimension $k$ \cite{costa2016random}.
In a similar vein to our model, random clique complexes \cite{kahle2009topology}, i.e., \SC{}s obtained by adding all cliques of a random graph as simplices, build on a graph as a 1-skeleton.
We refer to \cite{kahle2014topology,costa2016random} for a more detailed overview of existing models.
Existing variants of the configuration model for \SC{}s \cite{courtney2016generalized,young2017construction} focus on the connection between vertices and maximal simplices.
To the best of our knowledge, no bottom-up \SC{} model analogs to many of the well-known random graph models (e.g., modeling community structure, degree distribution, etc.) exist.

\subsection{Random (Abstract or Geometric) Cell Complex Models} 
On a theoretical level, random cell complexes in high-dimensional Euclidean spaces were introduced by Zähle \cite{zahle1988random}.
This results in geometric cell complexes, i.e., cell complexes that have an embedding and a distinction between cells that lie on the inside of the complex and those on the outside.
Moreover, to the best of our knowledge, there is no known efficient algorithm to sample from this distribution.
Further research on random geometric cell complexes \cite{schweinhart2016topological,leistritz1992topological} focused on tessellations of low- or high-dimensional point clouds and the local topology characterizing physical cell complexes.
Such geometric graphs (or \CC{}s) may accurately model some real-world datasets (e.g., communication between internet nodes) as they result in sparse, planar graphs with a possibly quite large diameter.
However, many applications have more dense, non-planar graphs with small diameters, such as telecommunication between cell towers \cite{barbarossa2020topological}, taxi trips (by endpoints) between neighborhoods \cite{taxi-dataset}, or social networks.
To date, researchers have filled the gaps in both models and sampling algorithms for existing models by using simple lifting procedures, lifting a well-structured graph to an abstract cell complex, or using a possibly biased sampling method.
Limiting the lifting to chordless cycles up to a certain length, for example, makes it computationally feasible to compute all possible 2-cells (in practice) \cite{bodnar2021weisfeiler}.
Alternatively, cells can be sampled from the Delaunay triangulation of a two-dimensional point cloud \cite{hoppe2023representing,sardellitti2021topological}.
However, these are ad hoc constructions to provide a useful lifting for certain datasets or synthetic data for an evaluation, respectively.
These methods have geometric properties and have not been studied in detail; they may therefore introduce biases into the resulting data: For example, \cite{hoppe2023representing} uses a process that consecutively augments triangles with neighboring triangles.
This growth process necessarily stops at the outer edges of the triangulation, making them more likely to be part of a cell boundary than inner edges.
The usage of these ad hoc models clearly indicates the need for models of random abstract cell complexes.

\end{document}